\documentclass{ws-rv9x6}             
\usepackage{amsmath,amsbsy,graphicx,indentfirst}

\newcommand{\R}{\mathbb{R}}

\renewcommand{\det}{{\rm Det}\,}
\newcommand{\gr}[1]{\boldsymbol{#1}}
\newcommand{\be}{\begin{equation}}
\newcommand{\ee}{\end{equation}}
\newcommand{\bea}{\begin{eqnarray}}
\newcommand{\eea}{\end{eqnarray}}
\newcommand{\ket}[1]{|#1\rangle}
\newcommand{\bra}[1]{\langle#1|}

\newcommand{\ketbra}[2]{\vert #1 \rangle \! \langle #2 \vert}
\newcommand{\N}{{\cal N}}

\newcommand{\sig}{\gr{\sigma}}
\newcommand{\bet}{\gr{\beta}}
\newcommand{\alp}{\gr{\alpha}}
\newcommand{\abs}[1]{\left\vert#1\right\vert}
\newcommand{\eq}[1]{Eq.~(\ref{#1})}

\begin{document}

\setcounter{chapter}{0}

\chapter{Bipartite and Multipartite
Entanglement of Gaussian States} \markboth{G. Adesso and F.
Illuminati}{Bipartite and Multipartite Entanglement of Gaussian
States}

\author{Gerardo Adesso and Fabrizio Illuminati}

\address{Dipartimento di Fisica ``E. R. Caianiello'',
Universit\`a di Salerno; CNR-Coherentia, Gruppo di Salerno; and INFN
Sezione di Napoli-Gruppo Collegato di Salerno, Via S. Allende,
84081 Baronissi (SA), Italy\\
E-mail: gerardo@sa.infn.it, illuminati@sa.infn.it}

\begin{abstract}
In this chapter we review the characterization of entanglement in
Gaussian states of continuous variable systems. For two-mode
Gaussian states, we discuss how their bipartite entanglement can be
accurately quantified in terms of the global and local amounts of
mixedness, and efficiently estimated by direct measurements of the
associated purities. For multimode Gaussian states endowed with
local symmetry with respect to a given bipartition, we show how the
multimode block entanglement can be completely and reversibly
localized onto a single pair of modes by local, unitary operations.
We then analyze the distribution of entanglement among multiple
parties  in multimode Gaussian states. We introduce the
continuous-variable tangle to quantify entanglement sharing in
Gaussian states and we prove that it satisfies the
Coffman-Kundu-Wootters monogamy inequality. Nevertheless, we show
that pure, symmetric three--mode Gaussian states, at variance with
their discrete-variable counterparts, allow a promiscuous sharing of
quantum correlations, exhibiting both maximum tripartite residual
entanglement and maximum couplewise entanglement between any pair of
modes. Finally, we investigate the connection between multipartite
entanglement and the optimal fidelity in a continuous-variable
quantum teleportation network. We show how the fidelity can be
maximized in terms of the best preparation of the shared entangled
resources and, viceversa, that this optimal fidelity provides a
clearcut operational interpretation of several measures of bipartite
and multipartite entanglement, including the entanglement of
formation, the localizable entanglement, and the continuous-variable
tangle.
\end{abstract}

\section{Introduction}

One of the main challenges in fundamental quantum theory as well as in quantum
information and computation sciences lies in the characterization and quantification
of bipartite entanglement for mixed states, and in the definition and interpretation
of multipartite entanglement both for pure states and in the presence of mixedness.
While important insights have been gained on these issues in the context of
qubit systems, a less satisfactory understanding has been achieved until
recent times on higher-dimensional systems, as the structure of entangled states
in Hilbert spaces of high dimensionality exhibits a formidable degree of complexity.
However, and quite remarkably, in infinite-dimensional Hilbert spaces of
continuous-variable systems, ongoing and coordinated efforts by different
research groups have led to important progresses in the understanding of the
entanglement properties of a restricted class of states, the so-called Gaussian
states. These states, besides being of great importance both from a fundamental
point of view and in practical applications, share peculiar features that make
their structural properties amenable to accurate and detailed theoretical analysis.
It is the aim of this chapter to review some of the most recent results on the
characterization and quantification of bipartite and multipartite entanglement
in Gaussian states of continuous variable systems, their relationships with standard
measures of purity and mixedness, and their operational interpretations in practical
applications such as quantum communication, information transfer, and quantum
teleportation.

\section{Gaussian States of Continuous Variable Systems}

We consider a continuous variable (CV) system consisting of $N$
canonical bosonic modes, associated to an infinite-dimensional
Hilbert space ${\cal H}$ and described by the vector $\hat{X}=\{\hat
x_1,\hat p_1,\ldots,\hat x_N,\hat p_N\}$ of the field quadrature
(``position'' and ``momentum'') operators. The quadrature phase
operators are connected to the annihilation $\hat a_i$ and creation
$\hat a^{\dag}_i$ operators of each mode, by the relations $\hat
x_{i}=(\hat a_{i}+\hat a^{\dag}_{i})$ and $\hat p_{i}=(\hat
a_{i}-\hat a^{\dag}_{i})/i$. The canonical commutation relations for
the $\hat X_i$'s can be expressed in matrix form: $[\hat X_{i},\hat
X_j]=2i\Omega_{ij}$, with the symplectic form
$\Omega=\oplus_{i=1}^{n}\omega$ and $\omega=\delta_{ij-1}-
\delta_{ij+1},\, i,j=1,2$.

Quantum states of paramount importance in CV systems are the
so-called Gaussian states, {\em i.e.}~states with Gaussian
characteristic functions and quasi--probability
distributions\cite{cvbook}. The interest in this special class of
states (important examples include vacua, coherent, squeezed,
thermal, and squeezed-thermal states of the electromagnetic field)
stems from the feasibility to produce and control them with linear
optical elements, and from the increasing number of efficient
proposals and successful experimental implementations of CV quantum
information and communication processes involving multimode Gaussian
states (see Ref.~\refcite{review} for recent reviews). By
definition, a Gaussian state is completely characterized by first
and second moments of the canonical operators. When addressing
physical properties invariant under local unitary transformations,
such as mixedness and entanglement, one can neglect first moments
and completely characterize Gaussian states by the $2N\times 2N$
real covariance matrix (CM) $\gr{\sigma}$, whose entries are
$\sigma_{ij}=1/2\langle\{\hat{X}_i,\hat{X}_j\}\rangle
-\langle\hat{X}_i\rangle\langle\hat{X}_j\rangle$. Throughout this
chapter, $\gr{\sigma}$ will be used indifferently to indicate the CM
of a Gaussian state or the state itself. A real, symmetric matrix
$\gr{\sigma}$ must fulfill the Robertson-Schr\"odinger uncertainty
relation\cite{simon87}
\begin{equation}\label{bonafide}
\gr{\sigma}+i\Omega \geq 0\,,
\end{equation}
to be a {\em bona fide} CM of a physical state. Symplectic
operations ({\em i.e.}~belonging to the group $Sp_{(2N,\R)}= \{S\in
SL(2N,\R)\,:\,S^T\Omega S=\Omega\}$) acting by congruence on CMs in
phase space, amount to unitary operations on density matrices in
Hilbert space. In phase space, any $N$-mode Gaussian state can be
transformed by symplectic operations in its Williamson diagonal
form\cite{williamson36}
 $\gr\nu$, such that $\gr{\sigma}= S^T \gr{\nu} S$,
with $\gr{\nu}=\,{\rm diag}\,\{\nu_1,\nu_1,\ldots\nu_N,\nu_N\}$. The
set $\Sigma=\{\nu_i\}$ of the positive-defined eigenvalues of
$|i\Omega\gr{\sigma}|$ constitutes the symplectic spectrum of
$\gr{\sigma}$ and its elements, the so-called symplectic
eigenvalues, must fulfill the conditions $\nu_i\ge 1$, following
from \eq{bonafide} and ensuring positivity of the density matrix
associated to $\gr{\sigma}$. We remark that the full saturation of
the uncertainty principle can only be achieved by pure $N$-mode
Gaussian states, for which $\nu_i=1\,\,\forall i=1,\ldots, N$.
Instead, those mixed states such that $\nu_{i\le k}=1$ and
$\nu_{i>k}>1$, with $1\le k\le N$, partially saturate the
uncertainty principle, with partial saturation becoming weaker with
decreasing $k$. The symplectic eigenvalues $\nu_i$ are determined by
$N$ symplectic invariants associated to the characteristic
polynomial of the matrix $|i\Omega\gr{\sigma}|$. Global
invariants include the determinant $\,{\rm
Det}\,\gr{\sigma}=\prod_{i}\nu_i^2$ and the quantity
 $\Delta=\sum_i \nu_i^2$, which is the sum of the
determinants of all the $2\times 2$ submatrices of $\gr{\sigma}$
related to each mode\cite{serafozzi}.

The degree of information about the preparation of a quantum state
$\varrho$ can be characterized by its \emph{purity} $\mu\equiv\,{\rm
Tr}\,\varrho^2$, ranging from $0$ (completely mixed states) to $1$
(pure states). For a Gaussian state with CM $\gr{\sigma}$ one
has\cite{paris} \be \label{muu} \mu=1/\sqrt{\,{\rm
Det}\,\gr{\sigma}}\,.\ee As for the entanglement, we recall that
positivity of the CM's partial transpose (PPT)\cite{ppt} is a
necessary and sufficient condition of separability for $(M+N)$-mode
bisymmetric Gaussian states (see Sec. 4) with respect to the $M|N$
bipartition of the modes\cite{unitarily}, as well as for
$(M+N)$-mode Gaussian states with fully degenerate symplectic
spectrum\cite{botero}. In the special, but important case $M=1$, PPT
is a necessary and sufficient condition for separability of all
Gaussian states\cite{simon00,wernerwolf}. For a general Gaussian
state of any $M|N$ bipartition, the PPT criterion is replaced by
another necessary and sufficient condition stating that a CM $\sig$
corresponds to a separable state if and only if there exists a pair
of CMs $\sig_{A}$ and $\sig_{B}$, relative to the subsystems $A$ and
$B$ respectively, such that the following inequality
holds\cite{wernerwolf}: $\sig \geq \sig_{A} \oplus \sig_{B}$. This
criterion is not very useful in practice. Alternatively, one can
introduce an operational criterion based on a nonlinear map, that is
independent of (and strictly stronger than) the PPT
condition\cite{giedkemappa}.

In phase space, partial transposition amounts to a mirror reflection
of one quadrature in the reduced CM of one of the parties. If
$\{\tilde{\nu}_i\}$ is the symplectic spectrum of the partially
transposed CM $\tilde{\gr{\sigma}}$, then a $(1+N)$-mode (or
bisymmetric $(M+N)$-mode) Gaussian state with CM $\gr{\sigma}$ is
separable if and only if $\tilde{\nu}_i\ge 1$ $\forall\, i$. A
proper measure of CV entanglement is the \emph{logarithmic
negativity}\cite{vidwer} $E_{\N}\equiv \log\|\tilde{\varrho}\|_{1}$,
where $\| \cdot \|_1$ denotes the trace norm, which constitutes an
upper bound to the {\em distillable entanglement} of the state
$\varrho$. It can be computed in terms of the symplectic spectrum
$\tilde{\nu}_i$ of $\tilde{\gr{\sigma}}$:
\begin{equation}\label{ensy}
E_{\N}= \max\left\{0,\,-{\sum}_{i :
\tilde{\nu}_i<1}\log\tilde{\nu}_i\right\}\,.
\end{equation}
$E_{\N}$ quantifies the extent to which the PPT condition
$\tilde{\nu}_i\ge 1$ is violated.

\section{Two--Mode Gaussian States: Entanglement and Mixedness}\label{secduemodi}
Two--mode Gaussian states represent the prototypical quantum states
of CV systems, and constitute an ideal test-ground for the
theoretical and experimental investigation of CV
entanglement\cite{francesi}. Their CM can be written is the
following block form
\begin{equation}
\boldsymbol{\sigma}\equiv\left(\begin{array}{cc}
\boldsymbol{\alpha}&\boldsymbol{\gamma}\\
\boldsymbol{\gamma}^{T}&\boldsymbol{\beta}
\end{array}\right)\, , \label{espre}
\end{equation}
where the three $2\times 2$ matrices $\boldsymbol{\alpha}$,
$\boldsymbol{\beta}$, $\boldsymbol{\gamma}$ are, respectively, the
CMs of the two reduced modes and the correlation matrix between
them. It is well known\cite{simon00} that for any two--mode CM
$\boldsymbol{\sigma}$ there exists a local symplectic operation
$S_{l}=S_{1}\oplus S_{2}$ which takes $\boldsymbol{\sigma}$ to its
standard form $\boldsymbol{\sigma}_{sf}$, characterized by
\begin{equation}
\gr\alpha={\rm diag}\{a,\,a\},\quad\gr\beta={\rm
diag}\{b,\,b\},\quad \gr\gamma={\rm diag}\{c_+,\,c_-\}\,.
\label{stform}
\end{equation}
States whose standard form fulfills $a=b$ are said to be
symmetric. Any pure state is symmetric and fulfills
$c_{+}=-c_{-}=\sqrt{a^2-1}$.  The uncertainty principle
Ineq.~(\ref{bonafide}) can be recast as a constraint on the
$Sp_{(4,{\mathbb R})}$ invariants ${\rm Det}\gr{\sigma}$ and
$\Delta(\gr{\sigma})={\rm Det}\boldsymbol{\alpha}+\,{\rm
Det}\boldsymbol{\beta}+2 \,{\rm Det}\boldsymbol{\gamma}$, yielding
$\Delta(\gr{\sigma})\le1+\,{\rm Det}\boldsymbol{\sigma}$.
The standard form covariances $a$, $b$, $c_{+}$, and $c_{-}$ can
be determined in terms of the two local symplectic invariants
\begin{equation}\label{mu12}
\mu_1 = (\det\gr\alpha)^{-1/2} = 1/a\,,\quad \mu_2 =
(\det\gr\beta)^{-1/2} = 1/b\,,
\end{equation}
which are the marginal purities of the reduced single--mode
states, and of the two global symplectic invariants
\begin{equation}\label{globinv}
\mu = (\det\gr\sigma)^{-1/2} =
[(ab-c_{+}^2)(ab-c_{-}^2)]^{-1/2}\,,\quad \Delta =
a^2+b^2+2c_+c_-\,,
\end{equation}
where $\mu$ is the global purity of the state.
Eqs.~(\ref{mu12}-\ref{globinv}) can be inverted to provide the
following physical parametrization of two--mode states in terms of
the four independent parameters $\mu_1,\,\mu_2,\,\mu$, and
$\Delta$\cite{extremal}:
\begin{equation}
a  \,\,=\,\, \frac{1}{\mu_1}\,, \quad b = \frac{1}{\mu_2}\,, \quad
c_{\pm}\,\,=\,\,\frac{\sqrt{\mu_1 \mu_2}}4 \, \big( \epsilon_- \pm
\epsilon_+ \big)\,, \label{gabc}
\end{equation}
with  $\epsilon_\mp \equiv \sqrt{\left[ \Delta - (\mu_1 \mp
\mu_2)^2/(\mu_1^2 \mu_2^2)\right]^2-4/\mu^2}$. The uncertainty
principle and the existence of the radicals appearing in \eq{gabc}
impose the following constraints on the four invariants in order to
describe a physical state
\begin{eqnarray}
\mu_1 \mu_2 &\le& \mu \,\,\le\,\,
\frac{\mu_1 \mu_2}{\mu_1 \mu_2 + \abs{\mu_1-\mu_2}}\,, \label{consmu} \\
\frac{2}{\mu} + \frac{(\mu_1 - \mu_2)^2}{\mu_1^2 \mu_2^2}
&\le& \Delta  \,\,\le\,\,  1+\frac{1}{\mu^2}  \, . \label{deltabnd}
\end{eqnarray}
The physical meaning of these constraints, and the role of the
extremal states ({\em i.e.~}states whose invariants saturate the
upper or lower bounds of Eqs.~(\ref{consmu}-\ref{deltabnd})) in
relation to the entanglement, will be investigated soon.

In terms of symplectic invariants, partial transposition
corresponds to flipping the sign of ${\rm Det}\,\gr{\gamma}$, so
that $\Delta$ turns into $\tilde{\Delta}=\Delta-4\,{\rm
Det}\,\gr{\gamma} = -\Delta + 2/\mu_1^2 + 2/\mu_2^2$. The
symplectic eigenvalues of the CM $\gr{\sigma}$ and of its partial
transpose $\tilde{\gr{\sigma}}$ are promptly determined in terms
of symplectic invariants
\begin{equation}
2\nu_{\mp}^2 = \Delta\mp\sqrt{\Delta^2
-4/\mu^2}\,, \quad
2\tilde{\nu}_{\mp}^2 = \tilde{\Delta}\mp\sqrt{\tilde{\Delta}^2
-4/\mu^2}\,, \label{n1}
\end{equation}
where in our naming convention $\nu_- \le \nu_+$ in general, and
similarly for the $\tilde\nu_\mp$. The PPT criterion yields a
state $\gr{\sigma}$ separable if and only if
$\tilde{\nu}_{-}\ge 1$.
Since $\tilde\nu_+ > 1$ for all two--mode Gaussian states, the
quantity $\tilde\nu_-$ also completely quantifies the entanglement,
in fact the logarithmic negativity \eq{ensy} is a monotonically
decreasing and convex function of $\tilde\nu_-$,
$E_{\N}=\max\{0,-\log\,\tilde{\nu}_{-}\}$. In the special instance
of symmetric Gaussian states, the \emph{entanglement of
formation}\cite{entfor} $E_F$ is also computable\cite{giedke03} but,
being again a decreasing function of $\tilde\nu_-$, it provides the
same characterization of entanglement and is thus fully equivalent
to $E_\N$ in this subcase.

A first natural question that arises is whether there can exist
two-mode Gaussian states of finite maximal entanglement at a given
amount of mixedness of the global  state. These states would be the
analog of the maximally entangled mixed states (MEMS) that are known
to exist for two-qubit systems\cite{hishizaka00}. Unfortunately, it
is easy to show that a similar question in the CV scenario is
meaningless. Indeed, for any fixed, finite global purity $\mu$ there
exist infinitely many Gaussian states which are infinitely
entangled. However, we can ask whether there exist maximally
entangled states at fixed global {\it and} local purities. While
this question does not yet have a satisfactory answer for two-qubit
systems, in the CV scenario it turns out to be quite interesting and
nontrivial. In this respect, a crucial observation is that, at fixed
$\mu$, $\mu_1$ and $\mu_2$, the lowest symplectic eigenvalue
$\tilde\nu_-$ of the partially transposed CM is a monotonically
increasing function of the global invariant $\Delta$. Due to the
existence of exact {\em a priori} lower and upper bounds on $\Delta$
at fixed purities (see Ineq.~\ref{deltabnd}), this entails the
existence of both maximally {\it and} minimally entangled Gaussian
states. These classes of extremal states have been introduced in
Ref.~\refcite{prl}, and completely characterized (providing also
schemes for their experimental production) in
Ref.~\refcite{extremal}, where the relationship between entanglement
and information has been extended considering generalized entropic
measures to quantify the degrees of mixedness. In particular, there
exist maximally and minimally entangled states also at fixed global
and local generalized Tsallis $p$-entropies\cite{extremal}. In this
short review chapter, we will discuss only the case in which the
purities (or, equivalently, the linear entropies) are used to
measure the degree of mixedness of a quantum state. In this
instance, the Gaussian maximally entangled mixed states (GMEMS) are
two--mode squeezed thermal states, characterized by a fully
degenerate symplectic spectrum; on the other hand, the Gaussian
least entangled mixed states (GLEMS) are states of partial minimum
uncertainty ({\em i.e.} with the lowest symplectic eigenvalue of
their CM being equal to 1). Studying the separability  of the
extremal states (via the PPT criterion), it is possible to classify
the entanglement properties of all two--mode Gaussian states in the
manifold spanned by the purities:
\begin{equation} \label{entsum}
\begin{array}{ll}
\mu_1 \mu_2 \; \le \; \mu \; \le \; \frac{\mu_1 \mu_2}{\mu_1
+ \mu_2 - \mu_1 \mu_2}, & \Rightarrow\ \hbox{separable;} \\
\frac{\mu_1 \mu_2}{\mu_1 + \mu_2 - \mu_1 \mu_2} < \mu \le
\frac{\mu_1 \mu_2}{\sqrt{\mu_1^2 + \mu_2^2 - \mu_1^2 \mu_2^2}},
& \Rightarrow\ \hbox{coexistence;} \\
\frac{\mu_1 \mu_2}{\sqrt{\mu_1^2 + \mu_2^2 - \mu_1^2 \mu_2^2}} <
\mu \le \frac{\mu_1 \mu_2}{\mu_1 \mu_2 + \abs{\mu_1-\mu_2}},
& \Rightarrow\ \hbox{entangled.} \\
\end{array}%
\end{equation}

In particular, apart from a narrow ``coexistence'' region where both
separable and entangled Gaussian states can be found, the
separability of two--mode states at given values of the purities is
completely characterized. For purities that saturate the upper bound
in Ineq.~(\ref{consmu}), GMEMS and GLEMS coincide and we have a
unique class of states whose entanglement depends {\it only} on the
marginal purities $\mu_{1,2}$. They are Gaussian maximally entangled
states for fixed marginals (GMEMMS). The maximal entanglement of a
Gaussian state decreases rapidly with increasing difference of
marginal purities, in analogy with finite-dimensional
systems\cite{adesso}. For symmetric states $(\mu_1=\mu_2)$ the upper
bound of Ineq.~(\ref{consmu}) reduces to the trivial bound $\mu \le
1$ and GMEMMS reduce to pure two--mode states.
\begin{figure}[t!]
\centering
\includegraphics[width=7.5cm]{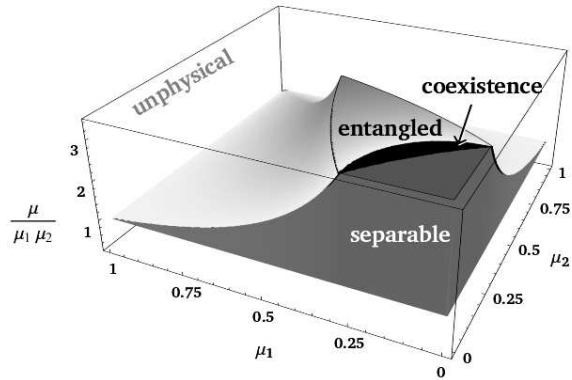}
\caption{ Classification of the entanglement for two--mode Gaussian
states in the space of marginal purities $\mu_{1,2}$ and normalized
global purity $\mu/\mu_1\mu_2$. All physical states lie between the
horizontal plane of product states $\mu=\mu_1\mu_2$, and the upper
limiting surface representing GMEMMS. Separable states (dark grey
area) and entangled states are well distinguished except for a
narrow coexistence region (depicted in black). In the entangled
region the average logarithmic negativity (see text) grows from
white to medium grey. The expressions of  the boundaries between all
these regions are collected in \eq{entsum}.} \label{fig2D}
\end{figure}
Knowledge of the global and marginal purities thus accurately
characterizes the entanglement of two-mode Gaussian states,
providing strong sufficient conditions and exact, analytical lower
and upper bounds. As we will now show, marginal and global purities
allow as well an accurate quantification of the entanglement.
Outside the region of separability, GMEMS attain maximum logarithmic
negativity $E_{{\N}\max}$  while, in the region of nonvanishing
entanglement (see Eq.~(\ref{entsum})), GLEMS acquire minimum
logarithmic negativity $E_{{\N}\min}$. Knowledge of the global
purity, of the two local purities, and of the global invariant
$\Delta$ ({\it i.e.}, knowledge of the full covariance matrix) would
allow for an exact quantification of the entanglement. However, we
will now show that an estimate based only on the knowledge of the
experimentally measurable global and marginal purities turns out to
be quite accurate. We can in fact quantify the entanglement of
Gaussian states with given global and marginal purities by the {\em
average logarithmic negativity} $\bar{E}_{\N} \equiv
(E_{{\N}\max}+E_{{\N}\min})/2$ We can then also define the relative
error $\delta \bar{E}_{\N}$ on $\bar{E}_{\N}$ as $\delta
\bar{E}_{\N} (\mu_{1,2},\mu) \equiv (E_{\N\max}
-E_{\N\min})/(E_{\N\max} +E_{\N\min})$. It is easy to see that this
error decreases {\em exponentially} both with increasing global
purity and decreasing marginal purities, {\it i.e.} with increasing
entanglement, falling for instance below  $5\%$ for symmetric states
($\mu_1=\mu_2\equiv\mu_i$) and $\mu > \mu_i$.
\begin{figure}[t!]
\centering
\begin{minipage}[t]{6.2cm}
\centering
\includegraphics[width=6.2cm]{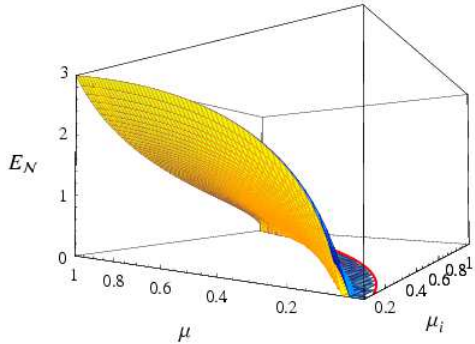}
\end{minipage} \hspace*{.3cm}
\begin{minipage}[b]{4.7cm}
\centering
\caption{Maximal and minimal logarithmic
negativities as functions of the global and marginal purities of
symmetric two-mode Gaussian states. The darker (lighter) surface represents GMEMS
(GLEMS). In this space, a generic two--mode mixed symmetric state
is represented by a dot lying inside the narrow gap between the
two extremal surfaces. \newline  \newline}
\label{fig3D}
\end{minipage}
\end{figure}
The reliable quantification of quantum correlations in genuinely
entangled two-mode Gaussian states is thus always assured by
the experimental determination of the purities, except at most for a
small set of states with very weak entanglement (states with $E_{\N}
\lesssim 1$). Moreover, the accuracy is even greater in the general
non-symmetric case $\mu_1 \neq \mu_2$, because the maximal
achievable entanglement decreases in such an instance. In
Fig.~\ref{fig3D}, the surfaces of extremal logarithmic negativities
are plotted versus $\mu_i$ and $\mu$ for symmetric states. In the
case $\mu=1$ the upper and lower bounds coincide, since
for pure states the entanglement is completely quantified by the
marginal purity. For mixed states this is not the case, but, as the
plot shows, knowledge of the global and marginal purities strictly
bounds the entanglement both from above and from below. This
analysis shows that the average logarithmic negativity
$\bar{E}_{\N}$ is a reliable estimate of the logarithmic negativity
$E_{\N}$, improving as the entanglement increases. We remark that the
purities may be directly measured experimentally, without the need
for a full tomographic reconstruction of the whole CM, by exploiting
quantum networks techniques\cite{network} or single--photon detections
without homodyning\cite{cerf}.

Finally, it is worth remarking that most of the results presented
here (including the sufficient conditions for entanglement based on
knowledge of the purities), being derived for CMs using the
symplectic formalism in phase space, retain their validity for
generic non Gaussian states of CV systems. For instance, any
two-mode state with a CM equal to that of an entangled two-mode
Gaussian state is entangled as well\cite{vanlok}. Our methods may
thus serve to detect entanglement for a broader class of states in
infinite-dimensional Hilbert spaces. The analysis briefly reviewed
in this paragraph on the relationships between entanglement and
mixedness, can be generalized to multimode Gaussian states endowed
with special symmetry under mode permutations, as we will show in
the next section.

\section{Multimode Gaussian States: Unitarily Localizable
Entanglement}\label{secuni}

We will now consider Gaussian states of CV systems with an arbitrary
number of modes, and briefly discuss the simplest instances
in which the techniques introduced for two--mode Gaussian states
can be generalized and turn out to be useful for the quantification
and the scaling analysis of CV multimode entanglement.
We introduce the notion of {\em bisymmetric} states, defined as those
$(M+N)$-mode Gaussian states, of a generic bipartition $M|N$, that
are invariant under local mode permutations
on the $M$-mode and $N$-mode subsystems. The CM $\sig$ of a
$(M+N)$-mode bisymmetric Gaussian state results from a correlated
combination of the fully symmetric blocks $\sig_{\alp^M}$ and
$\sig_{\bet^N}$: \be \sig = \left(\begin{array}{cc}
\sig_{\alp^M} & \gr\Gamma\\
\gr\Gamma^{\sf T} & \sig_{\bet^N}
\end{array}\right) \; , \label{fulsim}
\ee where $\sig_{\alp^M}$ ($\sig_{\bet^N}$) describes a $M$-mode
($N$-mode) reduced Gaussian state completely invariant under mode
permutations, and $\gr\Gamma$ is a $2M\times 2N$ real matrix formed
by identical $2\times 2$ blocks $\gr\gamma$. Clearly, $\gr\Gamma$ is
responsible for the correlations existing between the $M$-mode and
the $N$-mode parties. The identity of the submatrices $\gr\gamma$ is
a consequence of the local invariance under mode exchange, internal
to the $M$-mode and $N$-mode parties. A first observation is that
the symplectic spectrum of the CM $\sig$ \eq{fulsim} of a
bisymmetric $(M+N)$-mode Gaussian state includes two degenerate
eigenvalues, with multiplicities $M-1$ and $N-1$. Such eigenvalues
coincide, respectively, with the degenerate eigenvalue
$\nu_{\alpha}^{-}$ of the reduced CM $\sig_{\alpha^M}$ and the
degenerate eigenvalue $\nu_{\beta}^{-}$ of the reduced CM
$\sig_{\beta^N}$, with the same respective multiplicities. Equipped
with this result, one can prove\cite{unitarily} that $\sig$ can be
brought, by means of a local unitary operation, with respect to the
$M|N$ bipartition, to a tensor product of single-mode uncorrelated
states and of a two-mode Gaussian state with CM $\sig^{eq}$. Here we
give an intuitive sketch of the proof (the detailed proof is given
in Ref.~\refcite{unitarily}). Let us focus on the $N$-mode block
$\sig_{\beta^{N}}$. The matrices $i\gr{\Omega}\sig_{\beta^N}$ and
$i\gr{\Omega}\sig$ possess a set of $N-1$ simultaneous eigenvectors,
corresponding to the same (degenerate) eigenvalue. This fact
suggests that the phase-space modes corresponding to such
eigenvectors are the same for $\gr{\sigma}$ and for
$\gr{\sigma}_{\beta^{N}}$. Then, bringing by means of a local
symplectic operation the CM $\gr{\sigma}_{\beta^{N}}$ in Williamson
form, any $(2N-2)\times(2N-2)$ submatrix of $\gr{\sigma}$ will be
diagonalized because the normal modes are common to the global and
local CMs. In other words, no correlations between the $M$-mode
party with reduced CM $\sig_{\alpha^M}$ and such modes will be left:
all the correlations between the $M$-mode and $N$-mode parties will
be concentrated in the two conjugate quadratures of a single mode of
the $N$-mode block. Going through the same argument for the $M$-mode
block with CM $\sig_{\alpha^M}$ will prove the proposition and show
that the whole entanglement between the two multimode blocks can
always be concentrated in only two modes, one for each of the two
multimode parties.

While, as mentioned, the detailed proof of this result can be found
in Ref.~\refcite{unitarily} (extending the findings obtained in
Ref.~\refcite{adescaling} for the case $M=1$), here we will focus on
its relevant physical consequences, the main one being that the
bipartite $M\times N$ entanglement of bisymmetric $(M + N)$-mode
Gaussian states is {\em unitarily localizable}, {\em i.e.}, through
local unitary operations, it can be fully concentrated on a single
pair of modes, one belonging to party (block) $M$, the other
belonging to party (block) $N$. The notion of ``unitarily
localizable entanglement'' is different from that of ``localizable
entanglement'' originally introduced by Verstraete, Popp, and Cirac
for spin systems\cite{localiz}. There, it was defined as the maximal
entanglement concentrable on two chosen spins through local {\em
measurements} on all the other spins. Here, the local operations
that concentrate all the multimode entanglement on two modes are
{\em unitary} and involve the two chosen modes as well, as parts of
the respective blocks. Furthermore, the unitarily localizable
entanglement (when computable) is always stronger than the
localizable entanglement. In fact, if we consider a generic
bisymmetric multimode state of a $M|N$ bipartition, with each of the
two target modes owned respectively by one of the two parties
(blocks), then the ensemble of optimal local measurements on the
remaining (``assisting'') $M+N-2$ modes belongs to the set of local
operations and classical communication (LOCC) with respect to the
considered bipartition. By definition the entanglement cannot
increase under LOCC, which implies that the localized entanglement
(\`a la Verstraete, Popp, and Cirac) is always less or equal than
the original $M\times N$ block entanglement. On the contrary, {\it
all} of the same $M\times N$ original bipartite entanglement can be
unitarily localized onto the two target modes, resulting in a
reversible, of maximal efficiency, multimode/two-mode entanglement
switch. This fact can have a remarkable impact in the context of
quantum repeaters\cite{briegel} for communications with continuous
variables. The consequences of the unitary localizability are
manifold. In particular, as already previously mentioned, one can
prove that the PPT (positivity of the partial transpose) criterion
is a necessary and sufficient condition for the separability of $(M
+ N)$-mode bisymmetric Gaussian states\cite{unitarily}. Therefore,
the multimode block entanglement of bisymmetric (generally mixed)
Gaussian states with CM $\sig$, being equal to the bipartite
entanglement of the equivalent two-mode localized state with CM
$\sig^{eq}$, can be determined and quantified by the logarithmic
negativity in the general instance and, for all multimode states
whose two--mode equivalent Gaussian state is symmetric, the
entanglement of formation between the $M$-mode party and the
$N$-mode party can be computed exactly as well.

For the sake of illustration, let us consider fully symmetric
$2N$-mode Gaussian states described by a $2N \times 2N$ CM
$\sig_{\beta^{2N}}$. These states are trivially bisymmetric under
any bipartition of the modes, so that their block entanglement is
always localizable by means of local symplectic operations. This
class of states includes the pure, CV GHZ--type states (discussed in
Refs.\cite{vloock03,adescaling}) that, in the limit of infinite
squeezing, reduce to the simultaneous eigenstates of the relative
positions and the total momentum and coincide with the proper
Greenberger-Horne-Zeilinger\cite{ghzs} (GHZ) states of CV
systems\cite{vloock03}. The standard form CM $\sig^{p}_{\beta^{2N}}$
of this particular class of pure, symmetric multimode Gaussian
states depends only on the local mixedness parameter $b\equiv
1/\mu_\beta$, which is the inverse of the purity of any single-mode
reduced block, and it is proportional to the single-mode squeezing.
Exploiting our previous analysis, we can compute the entanglement
between a block of $K$ modes and the remaining $2N-K$ modes for pure
states (in this case the block entanglement is simply the Von
Neumann entropy of any of the reduced blocks) and, remarkably, for
mixed states as well.
\begin{figure}[t!]
\centering
\begin{minipage}[t]{6.7cm}
\centering
\includegraphics[width=6.7cm]{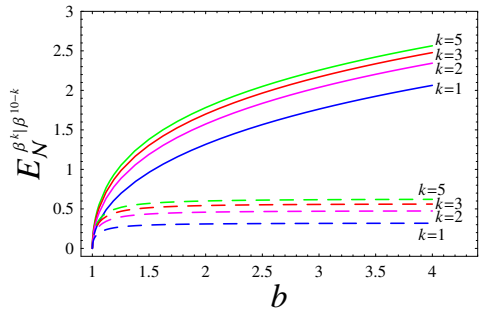}
\end{minipage} \hspace*{.3cm}
\begin{minipage}[b]{4.2cm}
\centering
\caption{Hierarchy of block entanglements of fully symmetric
$2N$-mode Gaussian states of $K \times (2N-K)$ bipartitions ($N=10$)
as a function of the single-mode mixedness $b$,  for pure states
(solid lines) and for mixed states obtained from $(2N+4)$-mode pure
states by tracing out $4$ modes (dashed lines). \newline \newline} \label{fiscalb}
\end{minipage}
\end{figure}

We can in fact consider a generic $2N$-mode fully symmetric mixed
state with CM $\sig_{\beta^{2N}}^{p\backslash Q}$, obtained from a
pure fully symmetric $(2N+Q)$-mode state by tracing out $Q$ modes.
For any $Q$ and any dimension $N$ of the block ($K \leq N$), and for
any nonzero squeezing ({\em i.e.~}for any $b>1$) one has that the
state exhibits genuine multipartite entanglement, as first remarked
in Ref.~\refcite{vloock03} for pure states: each $K$-mode party is
entangled with the remaining $(2N-K)$-mode block. Furthermore, the
genuine multipartite nature of the entanglement can be precisely
quantified by observing that the logarithmic negativity between the
$K$-mode and the remaining $(2N-K)$-mode block is an increasing
function of the integer $K \le N$, as shown in
Fig.~\ref{fiscalb}.The optimal splitting of the modes, which yields
the maximal, unitarily localizable entanglement, corresponds to
$K=N/2$ if $N$ is even, and $K=(N-1)/2$ if $N$ is odd. The multimode
entanglement of mixed states remains finite also in the limit of
infinite squeezing, while the multimode entanglement of pure states
diverges with respect to any bipartition, as shown in
Fig.~\ref{fiscalb}. For a fixed amount of local mixedness, the
scaling structure of the multimode entanglement with the number of
modes can be analyzed as well, giving rise to an interesting
result\cite{unitarily}. Let us consider, again for the sake of
illustration, the class of fully symmetric $2N$-mode Gaussian
states, but now at fixed single-mode purity. It is immediate to see
that the entanglement between any two modes decreases with $N$,
while the $N|N$ entanglement increases (and diverges for pure states
as $N \rightarrow \infty$): the quantum correlations become
distributed among all the modes. This is a clear signature of
genuine multipartite entanglement and suggests a detailed analysis
of its sharing properties, that will be discussed in the next
section. The scaling structure of multimode entanglement also
elucidates the power of the unitary localizability as a strategy for
entanglement purification, with its efficiency improving with
increasing number of modes. Finally, let us remark that the local
symplectic operations needed for the unitary localization can be
implemented by only using passive\cite{wolfito} and active linear
optical elements such as beam splitters, phase shifters and
squeezers, and that the original multimode entanglement can be
estimated by the knowledge of the global and local purities of the
equivalent, localized two--mode state (see
Refs.~\cite{adescaling,unitarily} for a thorough discussion), along
the lines presented in Section~\ref{secduemodi} above.

\section{Entanglement Sharing of Gaussian States}

Here we address the problem of entanglement sharing among multiple
parties, investigating the structure of multipartite
entanglement\cite{contangle,sharing}. Our aim is to analyze the
distribution of entanglement between different (partitions of) modes
in CV systems. In Ref.~\refcite{CKW} Coffman, Kundu and Wootters
(CKW) proved for a three-qubit system ABC, and conjectured for $N$
qubits (this conjecture has now been proven by Osborne and
Verstraete\cite{Osborne}), that the entanglement between, say, qubit
A and the remaining two--qubits partition (BC) is never smaller than
the sum of the A$|$B and A$|$C bipartite entanglements in the
reduced states. This statement quantifies the so-called {\em
monogamy} of quantum entanglement\cite{monogam}, in opposition to
the classical correlations which can be freely shared. One would
expect a similar inequality to hold for three--mode Gaussian states,
namely
\begin{equation}\label{CKWine}
E^{i|(jk)}- E^{i|j} - E^{i|k} \ge 0\,,
\end{equation}
where $E$ is a proper measure of CV entanglement and the indices
$\{i,j,k\}$ label the three modes. However, an immediate computation
on symmetric states shows that Ineq.~(\ref{CKWine}) can be violated
for small values of the single-mode mixedness $b$  using either the
logarithmic negativity $E_\N$ or the entanglement of formation $E_F$
to quantify the bipartite entanglement. This is not a
paradox\cite{sharing}; rather, it implies that none of these two
measures is the proper candidate for approaching the task of
quantifying entanglement sharing in CV systems. This situation is
reminiscent of the case of qubit systems, for which the CKW
inequality holds using the tangle $\tau$ \cite{CKW}, but fails if
one chooses equivalent measures of bipartite entanglement such as
the concurrence\cite{Wootters} ({\em i.e.}~the square root of the
tangle) or the entanglement of formation itself. Related problems on
inequivalent entanglement measures for the ordering of Gaussian
states are discussed in Ref.~\refcite{ordering}.

We then wish to define a new measure of CV entanglement able to
capture the entanglement distribution trade-off via the monogamy
inequality (\ref{CKWine}). A rigorous treatment of this problem is
presented in Ref.~\refcite{contangle}. Here we briefly review the
definition and main properties of the desired measure that
quantifies entanglement sharing in CV systems. Because it can be
regarded as the continuous-variable analogue of the tangle, we will
name it, in short, the {\it contangle}.

For a pure state $\ket{\psi}$ of a $(1+N)$-mode CV system,
we can formally define the contangle as
\begin{equation}\label{etaupure}
E_\tau (\psi) \equiv \log^2 \| \tilde \varrho \|_1\,,\quad \varrho =
\ketbra\psi\psi\,.
\end{equation}
$E_\tau (\psi)$ is a proper measure of bipartite entanglement, being
a convex, increasing function of the logarithmic negativity
$E_\N$, which is equivalent to the entropy of entanglement in
all pure states. For pure Gaussian states
$\ket\psi$ with CM $\sig^p$, one has $E_\tau (\sig^p) = \log^2 (1/\mu_1 -
\sqrt{1/\mu_1^2-1})$, where $\mu_1 = 1/\sqrt{\det\sig_1}$ is the
local purity of the reduced state of mode $1$, described by a CM
$\sig_1$ (considering $1 \times N$ bipartitions).
Definition (\ref{etaupure}) is extended to generic mixed
states $\varrho$ of $(N+1)$-mode CV systems through the convex-roof
formalism, namely:
\begin{equation}\label{etaumix}
E_\tau(\varrho) \equiv \inf_{\{p_i,\psi_i\}} \sum_i p_i
E_\tau(\psi_i)\,,
\end{equation}
where the infimum is taken over the decompositions of $\varrho$ in
terms of pure states $\{\ket{\psi_i}\}$. For infinite-dimensional
Hilbert spaces the index $i$ is continuous, the sum in \eq{etaumix}
is replaced by an integral, and the probabilities $\{p_i\}$ by a
distribution $\pi(\psi)$. All multimode mixed Gaussian states $\sig$
admit a decomposition in terms of an ensemble of pure Gaussian
states. The infimum of the average contangle, taken over all pure
Gaussian decompositions only, defines the {\em Gaussian contangle}
$G_{\tau}$, which is an upper bound to the true contangle $E_\tau$,
and an entanglement monotone under Gaussian local operations and
classical communications (GLOCC)\cite{ordering,geof}. The Gaussian
contangle, similarly to the Gaussian entanglement of
formation\cite{geof}, acquires the simple form $G_\tau (\sig) \equiv
\inf_{\sig^p \le \sig} E_\tau(\sig^p)$, where the infimum runs over
all pure Gaussian states  with CM $\sig^p \le \sig$.

Equipped with these properties and definitions, one can prove
several results\cite{contangle}. In particular, the general
(multimode) monogamy inequality $E^{i_m|(i_1\ldots i_{m-1}
i_{m+1}\ldots i_{N})} - \sum_{l \ne m} E^{i_m|i_l} \ge 0$  is
satisfied by all pure three-mode and all pure {\em symmetric}
$N$-mode Gaussian states, using either $E_\tau$ or $G_\tau$ to
quantify bipartite entanglement, and by all the corresponding mixed
states using $G_\tau$. Furthermore, there is numerical evidence
supporting the conjecture that the general CKW inequality should
hold for all {\it nonsymmetric} $N$-mode Gaussian states as
well.\footnote{The conjectured monogamy inequality for all (pure or
mixed) $N$-mode Gaussian states has been indeed proven by
considering a slightly different version of the continuous-variable
tangle, defined in terms of the (convex-roof extended) squared
negativity instead of the squared logarithmic negativity [T.
Hiroshima, G. Adesso and F. Illuminati, Phys. Rev. Lett. {\bf 98},
050503 (2007)].} The sharing constraint (\ref{CKWine}) leads to the
definition of the {\em residual contangle} as a tripartite
entanglement quantifier. For nonsymmetric three-mode Gaussian states
the residual contangle is partition-dependent. In this respect, a
proper quantification of tripartite entanglement is provided by the
{\em minimum} residual contangle
\begin{equation}\label{etaumin}
E_\tau^{i|j|k}\equiv\min_{(i,j,k)} \left[
E_\tau^{i|(jk)}-E_\tau^{i|j}-E_\tau^{i|k}\right]\,,
\end{equation}
where $(i,j,k)$ denotes all the permutations of the three mode
indexes. This definition ensures that $E_\tau^{i|j|k}$ is invariant
under mode permutations and is thus a genuine three-way property of
any three-mode Gaussian state. We can adopt an analogous definition
for the minimum residual Gaussian contangle $G_\tau^{i|j|k}$. One
finds that the latter is a proper measure of genuine tripartite CV
entanglement, because it is an entanglement
monotone under tripartite GLOCC for pure three-mode Gaussian
states\cite{contangle}.

Let us now analyze the sharing structure of multipartite CV
entanglement, by taking the residual contangle as a measure of
tripartite entanglement. We pose the problem of identifying
the three--mode analogues of the two inequivalent classes of fully
inseparable three--qubit states, the GHZ state\cite{ghzs}
$\ket{\psi_{\rm GHZ}} = (1/\sqrt{2}) \left[\ket{000} +
\ket{111}\right]$, and the $W$ state\cite{wstates} $\ket{\psi_{W}} =
(1/\sqrt{3}) \left[\ket{001} + \ket{010} + \ket{100}\right]$. These
states are both pure and fully symmetric, but the GHZ state
possesses maximal three-party tangle with no two-party quantum
correlations, while the $W$ state contains the maximal two-party
entanglement between any pair of qubits and its tripartite residual
tangle is consequently zero.

Surprisingly enough, in symmetric three--mode Gaussian states, if
one aims at maximizing (at given single--mode squeezing $b$) either
the two--mode contangle $E_\tau^{i|l}$ in any reduced state ({\it
i.e.}~aiming at the CV $W$-like state), or the genuine tripartite
contangle ({\it i.e.}~aiming at the CV GHZ-like state), one finds
the same, unique family of pure symmetric three--mode squeezed
states. These states, previously named ``GHZ-type''
states\cite{vloock03}, have been introduced for generic $N$--mode CV
systems in the previous Section, where their multimode entanglement
scaling has been studied\cite{adescaling,unitarily}. The peculiar
nature of entanglement sharing in this class of states, now baptized
CV GHZ/$W$ states, is further confirmed by the following
observation. If one requires maximization of the $1 \times 2$
bipartite contangle $E_\tau^{i|(jk)}$ under the constraint of
separability of all two--mode reductions, one finds a class of
symmetric mixed states whose tripartite residual contangle is
strictly smaller than the one of the GHZ/$W$ states, at fixed local
squeezing\cite{3modi}. Therefore, in symmetric three--mode Gaussian
states, when there is no two--mode entanglement, the three-party one
is not enhanced, but frustrated.

These results, unveiling a major difference between
discrete-variable and CV systems, establish the {\em promiscuous}
structure of entanglement sharing in symmetric Gaussian states.
Being associated with degrees of freedom with continuous spectra,
states of CV systems need not saturate the CKW inequality to achieve
maximum couplewise correlations. In fact, without violating the
monogamy inequality (\ref{CKWine}), pure symmetric three--mode
Gaussian states are maximally three-way entangled and, at the same
time, maximally robust against the loss of one of the modes due, for
instance, to decoherence, as demonstrated in full detail in
Ref.~\refcite{3modi}. This fact may promote these states,
experimentally realizable with the current technology\cite{3mexp},
as candidates for reliable CV quantum communication. Exploiting a
three--mode CV GHZ/$W$ state as a quantum channel can ensure for
instance a tripartite quantum information protocol like a
teleportation network or quantum secret sharing; or a standard,
highly entangled two--mode channel, after a unitary (reversible)
localization has been performed through a single beam splitter; or,
as well, a two--party quantum protocol with better-than-classical
efficiency, even if one of the modes is lost due to decoherence. We
will next focus on a relevant applicative setting of CV multipartite
entanglement, in which various of its properties discussed so far
will come in a natural relation.

\section{Exploiting Multipartite Entanglement: Optimal Fidelity of Continuous Variable Teleportation}

\begin{figure}[t!]
\centering
\begin{minipage}[t]{6.7cm}
\centering
\includegraphics[width=6.7cm]{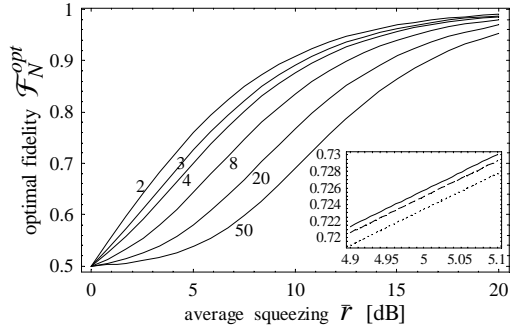}
 \end{minipage} \hspace*{.3cm}
\begin{minipage}[b]{4.2cm}
\centering  \caption{ Plot of the optimal fidelity for teleporting
an arbitrary coherent state from any sender to any receiver chosen
from $N$ ($N=2, \ldots, 50$) parties, exploiting $N$-party
entangled, pure symmetric Gaussian states as resources. A
nonclassical fidelity ${\cal F}^{opt}_N > 0.5$ is always assured for
any $N$, if the shared entangled resources are prepared in their
optimal form.
\newline}
\label{finopt}
\end{minipage}
\end{figure}

In this section we analyze an interesting application of
multipartite CV entanglement: a quantum teleportation-network
protocol, involving $N$ users who share a genuine $N$-partite
entangled Gaussian resource, completely symmetric under permutations
of the modes. In the standard multiuser protocol, proposed by Van
Loock and Braunstein\cite{bra00}, two parties are randomly chosen as
sender (Alice) and receiver (Bob), but, in order to accomplish
teleportation of an unknown coherent state, Bob needs the results of
$N-2$ momentum detections performed by the other cooperating
parties. A nonclassical teleportation fidelity (i.e. ${\cal F} >
{\cal F}^{cl}$) between {\em any} pair of parties is sufficient for
the presence of genuine $N$-partite entanglement in the shared
resource, while in general the converse is false (see {\it
e.g.}~Fig.~1 of Ref.~\refcite{bra00}). The {\em fidelity}, which
quantifies the success of a teleportation experiment, is defined as
${\cal F} \equiv \bra{\psi^{in}} \varrho^{out}\ket{\psi^{in}}$,
where ``in'' and ``out'' denote the input and the output state.
${\cal F}$ reaches unity only for a perfect state transfer,
$\varrho^{out} = \ket{\psi^{in}}\!\bra{\psi^{in}}$, while without
entanglement in the resource, by purely classical communication, an
average fidelity of ${\cal F}_{cl}=1/2$ is the best that can be
achieved if the alphabet of input states includes all coherent
states with even weight\cite{bfkjmo}. This teleportation network has
been recently  demonstrated experimentally\cite{naturusawa} by
exploiting three-mode squeezed Gaussian states\cite{3mexp}, yielding
a best fidelity of ${\cal F} = 0.64 \pm 0.02$, an index of genuine
tripartite entanglement. Our aim is to determine the optimal
multi-user teleportation fidelity, and to extract from it a
quantitative information on the multipartite entanglement in the
shared resources. By ``optimal'' here we mean maximization of the
fidelity over all local single-mode unitary operations, at fixed
amounts of noise and entanglement in the shared resource. We
consider realistically mixed $N$-mode Gaussian resources, obtained
by combining a mixed momentum-squeezed state (with squeezing
parameter $r_1$) and $N-1$ mixed position-squeezed states (with
squeezing parameter $r_2 \neq r_1$ and in principle a different
noise factor) into an $N$-splitter\cite{bra00} (a sequence of $N-1$
suitably tuned beam splitters). The resulting state is a completely
symmetric mixed Gaussian state of a $N$-mode CV system. For a given
thermal noise in the individual modes (comprising the unavoidable
experimental imperfections), all the states with equal average
squeezing $\bar r \equiv (r_1+r_2)/2$ are equivalent up to local
single--mode unitary operations and possess, by definition, the same
amount of multipartite entanglement with respect to any partition.
The teleportation efficiency, instead, depends separately on the
different single--mode squeezings. We have then the freedom of
unbalancing the local squeezings $r_1$ and $r_2$ without changing
the total entanglement in the resource, in order to single out the
optimal form of the resource state, which enables a teleportation
network with maximal fidelity. This analysis is straightforward (see
Ref.~\refcite{telepoppate} for details), but it yields several
surprising side results. In particular, one finds that the optimal
form of the shared $N$-mode symmetric Gaussian states, for $N>2$, is
neither unbiased in the $x_i$ and $p_i$ quadratures (like the states
discussed in Ref.~\refcite{bowen} for $N=3$), nor constructed by $N$
equal squeezers ($r_1=r_2= \bar r$). This latter case, which has
been implemented experimentally\cite{naturusawa} for $N=3$, is
clearly not optimal, yielding fidelities lower than $1/2$ for
$N\ge30$ and $\bar r$ falling in a certain interval\cite{bra00}.
According to the authors of Ref.~\refcite{bra00}, the explanation of
this paradoxical behavior should lie in the fact that their
teleportation scheme might not be optimal. However, a closer
analysis shows that the problem does not lie in the choice of the
protocol, but rather in the choice of the resource states. If the
shared $N$-mode squeezed states are prepared, by local unitary
operations, in the optimal form (described in detail in
Ref.~\refcite{telepoppate}), the teleportation fidelity ${\cal
F}^{opt}$ is guaranteed to be nonclassical (see Fig.~\ref{finopt})
as soon as $\bar r>0$ for any $N$, in which case the considered
class of pure states is genuinely multiparty entangled, as we have
shown in the previous sections. In fact, one can
show\cite{telepoppate} that this nonclassical optimal fidelity is
{\em necessary and sufficient} for the presence of multipartite
entanglement in any multimode symmetric Gaussian state used as a
shared resource for CV teleportation.
These findings yield quite naturally a direct operative way to
quantify multipartite entanglement in $N$-mode (mixed) symmetric
Gaussian states, in terms of the so-called {\em Entanglement of
Teleportation}\cite{telepoppate}, defined as the normalized optimal
fidelity
\begin{equation}\label{et}
E_T \equiv \max\left\{0, \big({\cal F}_N^{opt}-{\cal F}_{cl}\big)/
\big(1-{\cal F}_{cl}\big)\right\}\,,
\end{equation}
going from 0 (separable states) to 1 (CV GHZ/$W$ state). Moreover,
one finds that the optimal shared entanglement that allows for the
maximal fidelity is {\it exactly} the CV counterpart of the
localizable entanglement, originally introduced for spin systems by
Verstraete, Popp, and Cirac\cite{localiz}. The CV localizable
entanglement (not to be confused with the unitarily localizable
entanglement introduced in Section \ref{secuni}) thus acquires a
suggestive operational meaning in terms of teleportation processes.
In fact, the localizable entanglement of formation (computed by
finding the optimal set of local measurements
--- unitary transformations and nonunitary momentum detections ---
performed on the assisting modes to concentrate the highest possible
entanglement onto Alice and Bob pair of modes) is a monotonically
increasing function of $E_T$: $E_F^{loc} = f[(1-E_T)/(1+E_T)]$, with
$f(x) \equiv \frac{(1+x)^2}{4x} \log{\frac{(1+x)^2}{4x}} -
\frac{(1-x)^2}{4x} \log{\frac{(1-x)^2}{4x}}$. For $N=2$ (standard
two-user teleportation\cite{brakim}) the state is already localized
and $E_F^{loc} = E_F$, so that $E_T$ is equivalent to the
entanglement of formation $E_F$ of two-mode Gaussian states.
Remarkably, for $N=3$, {\it i.e.~} for three-mode pure Gaussian
resource states, the residual contangle $E_\tau^{i|j|k}$ introduced
in Section 5 (see \eq{etaumin}) turns out to be itself a
monotonically increasing function of $E_T$:
\begin{equation}
E_\tau^{i|j|k} = \log^2{\frac{2\sqrt2
E_T-(E_T+1)\sqrt{E_T^2+1}}{(E_T-1)\sqrt{E_T(E_T+4)+1}}}-\frac12
\log^2\!{\frac{E_T^2+1}{E_T(E_T+4)+1}}\,.
\end{equation}
The quantity $E_T$ thus represents another {\em
equivalent} quantification of genuine tripartite CV
entanglement and provides the latter with an
operational interpretation associated to the success of a
three-party teleportation network. This suggests a possible
experimental test of the promiscuous sharing of CV entanglement,
consisting in the successful (with nonclassical optimal fidelity)
implementation of both a three-user teleportation network exploiting
pure symmetric Gaussian resources, and of two-user standard
teleportation exploiting any reduced two-mode channel obtained
discarding a mode from the original resource.

Besides their theoretical aspects, the results
reviewed in this section are of direct practical
interest, as they answer the experimental need for the best
preparation recipe of an entangled squeezed resource, in order to
implement quantum teleportation and in general CV communication
schemes with the highest possible efficiency.

\section{Conclusions and Outlook}

We have reviewed some recent results on the entanglement of Gaussian
states of CV systems. For two-mode Gaussian states we have shown how
bipartite entanglement can be qualified and quantified via the
global and local degrees of purity. Suitable generalizations of the
techniques introduced for two-mode Gaussian states allow to analyze
various aspects of entanglement in multimode CV systems, and we have
discussed recent findings on the scaling, localization, and sharing
properties of multipartite entanglement in symmetric, bisymmetric,
and generic multimode Gaussian states. Finally, we have shown that
many of these properties acquire a clear and simple operational
meaning in the context of CV quantum communication and teleportation
networks. Generalizations and extensions of these results appear at
hand, and we may expect further progress along these lines in the
near future, both for Gaussian and non Gaussian states.

A good portion of the material reported in this chapter originates
from joint work with our friend and colleague Alessio Serafini, whom
we warmly thank for the joy of collaborating together. It is as well
a pleasure to acknowledge stimulating exchanges over the last two
years with Nicolas Cerf, Ignacio Cirac, Jens Eisert, Jarom\'ir
Fiur\'a\u sek, Ole Kr\"uger, Gerd Leuchs, Klaus M{\o}lmer, Tobias Osborne, Matteo
Paris, Eugene Polzik, Gustavo Rigolin, Peter van Loock, Frank
Verstraete, David Vitali, Reinhard Werner, Michael Wolf,
and Bill Wootters.

\end{document}